\documentclass[twocolumn,aps,prl]{revtex4-1}

	\usepackage[usenames,dvipsnames]{xcolor}
	\usepackage{amsmath}
	\usepackage{amsfonts}
	\usepackage{amssymb}
	\usepackage[colorlinks=true,citecolor=blue,linkcolor=blue,urlcolor=blue]{hyperref}
	\usepackage{graphicx}
	\usepackage{bbold}					
	\usepackage[makeroom]{cancel}		
	\usepackage{multirow}				
	\usepackage[normalem]{ulem}        
	\usepackage{array}
	\usepackage{pdfpages}
    \usepackage{float}

	\makeatletter
	\AtBeginDocument{\let\LS@rot\@undefined}
	\makeatother
	
	\newcolumntype{x}[1]{>{\centering\let\newline\\\arraybackslash\hspace{0pt}}p{#1}}

	\DeclareMathAlphabet{\mathbbold}{U}{bbold}{m}{n}

	\newcounter{subeqn} %

	\makeatletter
	\@addtoreset{subeqn}{equation}
	\makeatother

\definecolor{TB}{rgb}{0,0,0} 

\def\Eq#1{Eq.~(\ref{#1})}
\def\Fig#1{Fig.~\ref{#1}}

\begin{document}

\title{Order-by-disorder and emergent Kosterlitz-Thouless phase in triangular Rydberg array}

\author{Sibo Guo$^{1,2}$}
\author{Jiangping Hu$^{1,2,3}$}\email[Corresponding author: ]{jphu@iphy.ac.cn}
\author{Zi-Xiang Li$^{1,2}$}\email[Corresponding author: ]{zixiangli@iphy.ac.cn}

\affiliation{$^{1}$Beijing National Laboratory for Condensed Matter Physics,
	and Institute of Physics, Chinese Academy of Sciences, Beijing 100190, China}
\affiliation{$^{2}$School of Physical Sciences, University of Chinese Academy of Sciences, Beijing 100049, China}
\affiliation{$^{3}$Kavli Institute for Theoretical Sciences, University of Chinese Academy of Sciences, Beijing 100190, China}

\date{\today}

\begin{abstract}
Programmable quantum simulator using Rydberg-atom array provides a promising route to demystifying quantum many-body physics in strongly correlated systems. Motivated by recent realization of various quantum magnetic phases on frustrated Rydberg-atom array, we perform numerically exact quantum Monte-Carlo simulation to investigate the exotic states of matter emerging in the model describing Rydberg atom on triangular lattice. Our state-of-the-art simulation unveils the $\sqrt{3}\times\sqrt{3}$ triangular anti-ferromagnetic(TAF) order exists at $\frac{1}{3}$/$\frac{2}{3}$-Rydbreg filling, consistent with the observation in experiments. More fantastically, $\sqrt{3}\times\sqrt{3}$ long-range order arising from order-by-disorder mechanism emerges at $\frac{1}{2}$-filling. At finite temperature, $U(1)$ symmetry is emergent at $\frac{1}{2}$-filling and a Kosterlitz-Thouless(KT) phase transition occurs with increasing temperature. These intriguing phenomena are potentially detected in future Rydberg-atom experiments.
\end{abstract}

\maketitle

{\em Introduction.}---Deciphering the exotic physics emerging in frustrated magnets is a particularly intriguing and crucial subject in strongly correlated physics, triggering enduring interests in many years\cite{BalentsReview,ZhouReview}. In light of recent experimental developments, the models featuring quantum magnetism in the presence of lattice frustration are realized in Rydberg atom arrays\cite{Browaeys2021Nature,Lukin2021Nature,Lukin2021Science}. In virtual of the high controllability and precise measurability in Rydberg atom set-up\cite{r203,r204,r205,r206,r207,r208,r209,r210,r188,r213,r216,r218,r219,r222}, fruitful exotic phenomena emerging in frustrated magnets, including various spin density wave orders and quantum spin liquid, have been observed in the platform of Rydberg atom array\cite{r174,r177,r169,Lukin2021Science,Lukin2017Nature,r170,r166,r168,r164,r178,r176,Lukin2021Nature,Browaeys2021Nature}. Hence, Rydberg atom array provides an ideal candidate platform for quantum simulation on quantum magnetism\cite{Vishwanath2021PRX,Samajdar2022arXiv,Choi2022PRB,Zhang2022PRB,r164,r172,r173,r183,r185,r189,r190,r192,r194,r195,
r211,r212,r214,r215,r220,r227,r238,r240,r241,r247,r248,r255,r259,yan2023emergent} and other exotic phenomena in strongly correlated systems\cite{Lukin2019Nature,Browaeys2019Science,Serbyn2018NP,Xu2022PRL,Browaeys2020PRX,Zhai2022arXiv,Bakr2021PRX,ohler2022arXiv,Zhu2022arXiv,Turner2021PRX,Liu2022arXiv,r179,r223,r229,r230,r235,r239,r246,r249,r252,r257}.

Triangular Ising model with transverse field is a representative model featuring quantum magnetism with lattice frustration. Despite its simple form, triangular transverse Ising model is theoretically revealed to exhibit various intriguing phenomena arising from the interplay between lattice frustration and quantum/thermal fluctuation. For example, quantum fluctuation triggers a $\sqrt{3}\times\sqrt{3}$ ordered phase, emerging from the macroscopic degenerate states induced by the lattice frustration, which is dubbed as order-by-disorder(OBD) mechanism\cite{Henley1989PRL,Chubukov1992PRB,Chubukov1992PRL,Moessner2000PRL,Moessner2000PRB,Kim2008PRL,Trebst2014PRB,Cai2021PRB,OBOReview,Balents2005PRL,Chen2020PRR}. At finite temperature, there exists an intermediate critical phase with emergent $U(1)$ symmetry, namely KT phase, owing to the interplay of quantum and thermal fluctuations in the presence of lattice frustration \cite{Sandvik2003PRE,Damle2015PRL,Yu2020NC,Li2020NC,Wan2020PRB,Meng2021PRB,Dun2021PRB}. Recently, Rydberg atom array with up to 196 atoms and high coherence on triangular lattice is realized experimentally, featuring Ising-like model involving long-range interactions\cite{Browaeys2021Nature}. Remarkably, the classical pattern of $\sqrt{3}\times\sqrt{3}$ AFM order is observed. Nevertheless, the realization of OBD phase and associated intriguing physics such as emergent $U(1)$ symmetry in the Rydberg atoms platform remains elusive. A thorough study of the systems with numerically exact theoretical approach, rigorously illustrating the plausibility of realization of the exotic physics discussed above in triangular Rydberg atom array, is thus immensely desired.

\begin{figure}[t]
	\centerline{\includegraphics[scale=0.13]{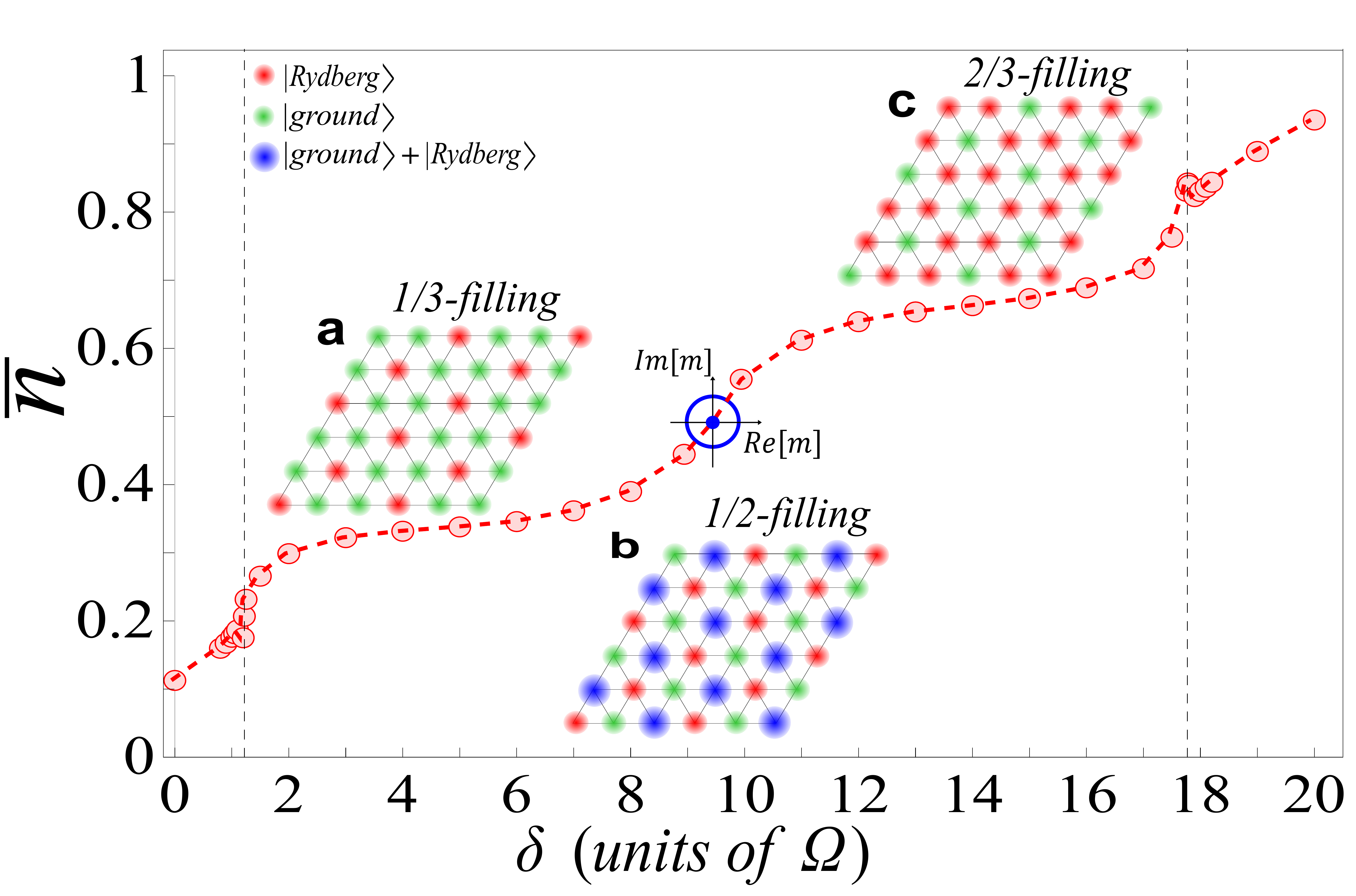}}
	\caption{{\bf Average Rydberg filling.} The red dots connected by the dashed line represent the Rydberg filling versus $\delta$ in model \Eq{Eq1}. Insets (a)-(c) show the patterns of TAF order at $\frac{1}{3}$-, $\frac{1}{2}$- and $\frac{2}{3}$-filling, respectively. The blue circle represents the distribution of TAF order parameters at $1/2$-filling, which indicates the appearance of $U(1)$ symmetry. }
	\label{Fig1}
\end{figure}

In this paper, we perform numerically exact quantum Monte-Carlo(QMC) simulation based on stochastic series expansion (SSE)\cite{Sandvik1999PRB,Sandvik2003PRE,r161} algorithm to investigate a realistic model describing Rydberg atom array on triangular lattice\cite{Browaeys2021Nature}. We implement the algorithm combining line update \cite{r159} and multi-branch update, enabling studying the model at low temperature and large system size with relatively high efficiency. Our state-of-the-art QMC simulation reveals the existence of $\sqrt{3}\times\sqrt{3}$ TAF order at $\frac{1}{3}$- and $\frac{2}{3}$-filling, consistent with the observation in experiments. More remarkably, our study demonstrates the emergence of OBD driven $\sqrt{3}\times\sqrt{3}$ ordered phase at $\frac{1}{2}$-filling. With increasing temperature, the TAF ordered phases at $\frac{1}{3}$- and $\frac{2}{3}$-filling melt into disordered phase through a continuous transition belonging to two-dimensional Potts universality class. Intriguingly, at $\frac{1}{2}$-filling, the $U(1)$ isotropy emerges in a large temperature regime and the associated transition from the OBD phase to high-temperature disordered phase thus belongs to the KT universality class\cite{KTpaper}. Our study paves a novel avenue to theoretically investigate the Rydberg atom array system in a numerically exact approach. The results provide concrete theoretical support to potentially realizing the exotic phenomena, for instance OBD phase, emergent $U(1)$ symmetry and KT phase transition, in Rydberg atom platform.

{\em Model and method.}--- We consider the following Hamiltonian describing 2D interacting Rydberg array~\cite{Browaeys2021Nature}:
\begin{equation}
\begin{aligned}
\hat{H}=\sum_{i<j}U_{ij}\hat{n}_{i}\hat{n}_{j}+\frac{\hbar\Omega}{2}\sum_{i}(\hat{b}^\dagger_i + \hat{b}_i) -\hbar \delta \sum_{i}\hat{n} _{i},\\
	\label{Eq1}
\end{aligned}
\end{equation}
where $\hbar$ is reduced Planck's constant, $\Omega$ is Rabi frequency, $\delta$ is detuning frequency and $\hat{n}_i$ denotes Rybderg occupation on site $i$. The model features a long-range van der Waals interaction between Rydberg atoms $U_{ij} = \Omega(R_{b}/r_{ij})^{6}$, where $R_b$ is Rydberg blocking radius defined as $U_{ij}=\Omega$ if $r_{ij}=R_b$. We investigate \Eq{Eq1} on triangular lattice with the lattice basis vectors $\vec{a}_1 = a(1,0)$ and $\vec{a}_2 = a(\frac{1}{2},\frac{\sqrt{3}}{2})$. Unless specified, we choose $\hbar=1$ and $\Omega=1$ in the simulation.

Upon fixing $R_b$, with the increase of detuning frequency $\delta$, atoms in the ground state are gradually excited to the Rydberg state, exhibiting various patterns in real space with different average Rydberg fillings\cite{r172,r173,Browaeys2021Nature}. In particular, as blockade radius is comparable to lattice constant on triangular lattice, namely $a \leq R_{b} \leq \sqrt{3}a$, the Rydberg atoms form an incompressible TAF pattern with $\sqrt{3}\times\sqrt{3}$ periodicity as revealed in Ref.\cite{Browaeys2021Nature}. In this study, we fix the blockade radius $R_b=1.2a$ and select the system size $L_x=L_y=L$ with periodic boundary condition.
Owing to the absence of sign problem\cite{r159,Li2019Review,Li2016PRL,Xiang2016PRL,Wang2015PRL,Li2015PRB}, the accurate properties of \Eq{Eq1} for large system size and low temperature are accessible in QMC simulation.
The results of other choices of $R_b$ such as $R_b=1.0a$ and the details of QMC simulation are included in supplemental material (SM)\cite{rSM3}.

\begin{figure}[t]
	\centerline{\includegraphics[scale=0.3]{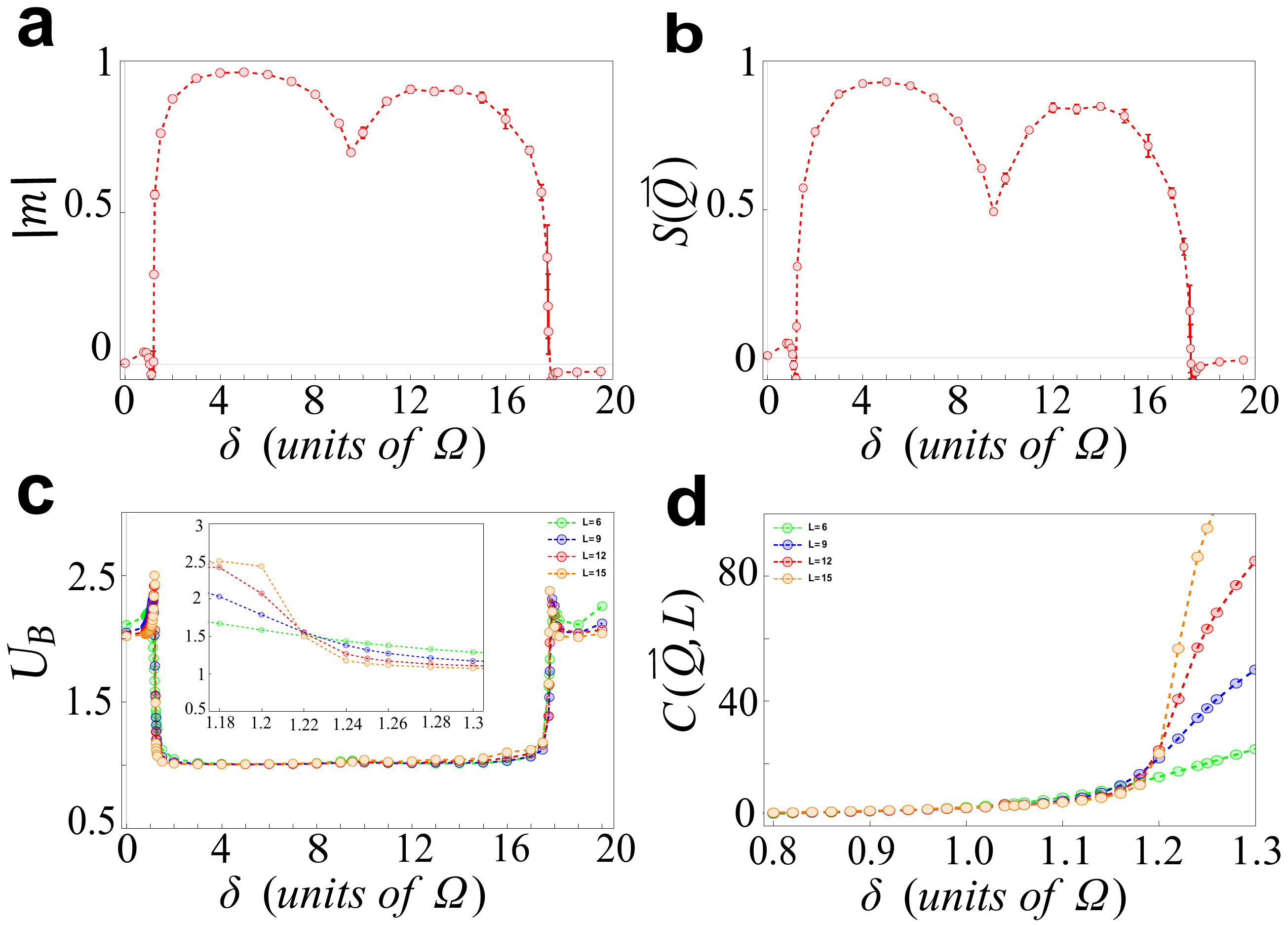}}
	\caption{{\bf Ground-state Phase diagram and transition point.} (a) and (b) represent the results of order parameter and static structure factor versus $\delta$ at thermodynamic limit, respectively.  (c) Binder ratio versus $\delta$ for $L=6,9,12,15$. The inset provides a zoom-in view of $U_{B}$ in the regime close to the transition point. (d) Correlation length ratio versus $\delta$ in the regime close to the transition point. }
	\label{Fig2}
\end{figure}

{\em Phase diagram.}---We calculate the average Rydberg filling $\bar{n}$ at zero temperature versus $\delta$ as depicted in Fig.~\ref{Fig1}, showing the Rydberg filling gradually increases from $0.1$ to $1$ with increasing $\delta$. The results explicitly indicate two plateaus appear at $\frac{1}{3}$- and $\frac{2}{3}$-filling associated with incompressible phases. To further reveal the spontaneous symmetry breaking (SSB) patterns of Rydberg atoms, we calculate the order parameter $m$ and the static structure factor $S(\vec{Q})$ corresponding to the $\sqrt{3} \times \sqrt{3}$ TAF phase, defined as the Fourier transform of the Rydberg occupation $\hat{n}_{i}$ and the correlation function between $\hat{n}_{i}$ and $\hat{n}_{j}$, respectively,
\begin{equation}
\begin{aligned}
 m= \sum_{i}\left<\hat{n}_{i} \right>e^{i\vec{Q}\cdot \vec{x}_{i}}/(N/3)
 	\label{Eq2}
\end{aligned}
\end{equation}
and
\begin{equation}
\begin{aligned}
 S(\vec{Q}) =\frac{1}{N}\sum_{i,j}e^{i\vec{Q}\cdot (\vec{x}_{i}-\vec{x}_{j})}\left<\hat{n} _{i}\hat{n} _{j}\right>.
 	\label{Eq3}
\end{aligned}
\end{equation}
Here, $\vec{Q}=\frac{1}{a}(\frac{2\pi}{3},\frac{2\sqrt{3}\pi}{3})$ and $N$ is the total number of lattice points. $m$ and $S(\vec{Q})$ at thermodynamic limit ($L\rightarrow \infty$) are extracted from results of $L=3\sim18$, the details of which are included in SM\cite{rSM3}. As depicted in Fig.~\ref{Fig2} (a) and (b),  $|m|$ and $S(\vec{Q})$ are finite at thermodynamic limit in a large regime of $\delta$, indicating the existence of $\sqrt{3}\times\sqrt{3}$ TAF long-range order. Both of these quantities exhibit the feature of pronounced convex in the regime corresponding to the $\frac{1}{3}$/$\frac{2}{3}$-filling platform in \Fig{Fig1},
suggesting the ground state of \Eq{Eq1} possesses strong $\sqrt{3}\times\sqrt{3}$ TAF long-range order at the commensurate $\frac{1}{3}$- and $\frac{2}{3}$-filling, consistent with the observation in Rydberg atom array experiments\cite{Browaeys2021Nature}. With increasing $\delta$, the results of $|m|$ and $S(\vec{Q})$ display sharp discontinuous transitions around $\delta_c \approx 1.2/17.8$, signaling the nature of first-order transition between disordered and TAF phases.

To further corroborate the emergence of $\sqrt{3}\times\sqrt{3}$ TAF long-range order and determine the accurate transition point, we perform finite-size scaling analysis by calculating the Binder ratio\cite{r163,r153,r155,r172} defined as $U_{B}(L)= \frac{\left< |m|^{4}\right>}{\left< |m|^{2}\right>^{2}}$. In the long (short)-range ordered phase, the values of $U_{B}(L)$ should decrease (increase) with increasing system size $L$ and tend to 1 (2) in the thermodynamic limit. Fig.~\ref{Fig2} (c) shows the results of $U_{B}$ versus $\delta$ and the nearly crossing point from different sizes $L$ indicates the phase transition occurring at $\delta/\Omega=1.22\pm 0.02$, in agreement with the results of $|m|$ and $S(\vec{Q})$. The non-monotonic behavior of $U_{B}(L)$ close to the phase transition point manifests the nature of first-order transition. Additionally, we calculate the correlation length ratio \cite{r161,r173} defined as: $C(\vec{Q},L)= \frac{S(\vec{Q})}{S(\vec{Q}+\vec{\delta}_q)}$,
where $\vec{Q}+\vec{\delta}_q$ is the momentum closest to $\vec{Q}$ on the lattice. As depicted in \Fig{Fig2}(d), the results of $C(\vec{Q},L)$ for different system sizes reveal the transition from disordered to TAF phase occurs at $\delta_{c1}\approx 1.2$. Similarly, $U_{B}(L)$ and $C(\vec{Q},L)$ yield another phase boundary between TAF and disordered phases $\delta_{c2} \approx 17.8$. Taken together, our systematic numerical simulations utilizing various approaches unambiguously establish the ground-state phase diagram of Rydberg atoms on triangular lattice as presented in \Fig{Fig1} and \Fig{Fig2}. Notice that the ground-state phase diagram does not qualitatively change with respect to slightly varying the value of $R_b$, as suggested by the numerical results of $R_b=1.0$ included in SM\cite{rSM3}.

\begin{figure}[t]
	\centerline{\includegraphics[scale=0.13]{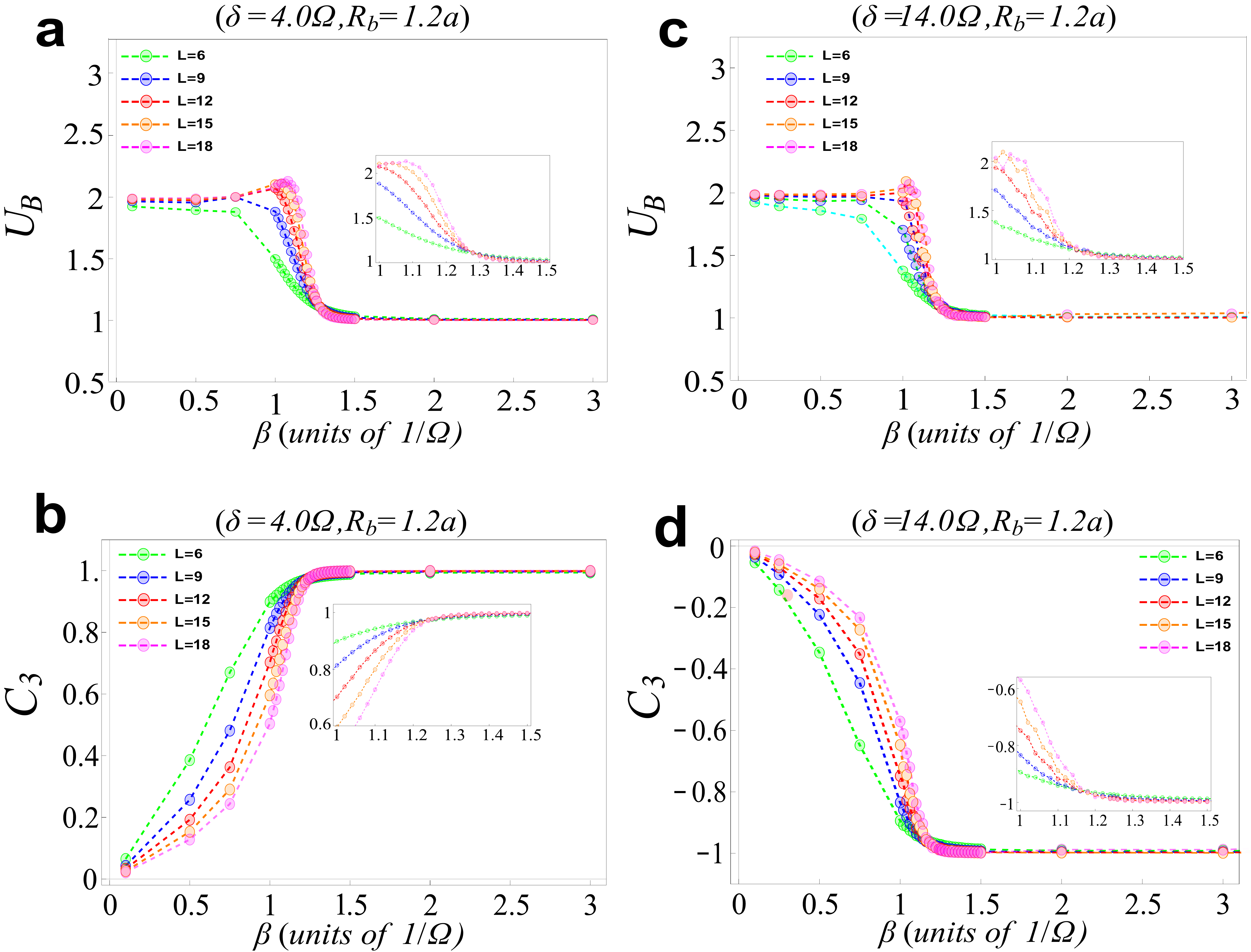}}
	\caption{{\bf Transition of TAF phase at finite temperature.} (a) and (c) represent the results for Binder ratio as a function of $\beta$ at $\frac{1}{3}$- and $\frac{2}{3}$-filling, respectively. (b) and (d) represent the results for $C_{3}$ as a function of $\beta$ at $\frac{1}{3}$- and $\frac{2}{3}$-filling, respectively. The system sizes in the simulation are $L=6,9,12,15,18$. The insets provide a zoom-in view in the regime close to the transition point.}
	\label{Fig3}
\end{figure}

{\em OBD at $1/2$-filling.}---The $\sqrt{3}\times\sqrt{3}$ TAF phases at $\frac{1}{3}$/$\frac{2}{3}$-filling established in our state-of-the-art numerical simulation are consistent with the recent experimental observations\cite{Browaeys2021Nature}. An intriguing question is whether the $\sqrt{3}\times\sqrt{3}$ TAF order could emerge at $\frac{1}{2}$-filling, arising from the celebrated OBD mechanism as developed in triangular quantum transverse-field Ising model\cite{Henley1989PRL,Chubukov1992PRB,Chubukov1992PRL,Moessner2000PRL,Moessner2000PRB}. In Fig.~\ref{Fig2} (a) and (b), the finite order parameter and structure factors at thermodynamic limit demonstrate the presence of $\sqrt{3}\times\sqrt{3}$ TAF long-range order at $\frac{1}{2}$-filling. Furthermore, the results of Binder ratio in Fig.~\ref{Fig2} (c) rigorously confirm the conclusion. The explicit results of order parameters, structure factors, Binder/correlation length ratios at $\frac{1}{2}$-filling for different system sizes are included in SM\cite{rSM3}.

{\em KT phase with $U(1)$ symmetry.}---From the perspective of symmetry, the TAF phase at $\frac{1}{3}$/$\frac{2}{3}$-filling breaks $Z_3$ lattice translational (or rotational) symmetry. At $\frac{1}{2}$-filling, the system accommodates an extra $Z_2$ particle-hole symmetry, namely the ground state of TAF phase is six-fold degenerate. Hence, the resulting Landau free energy characterizing TAF ordering is written as:
\begin{equation}
\begin{aligned}
F\propto & ~g_{2}|m|^2+g_{4}|m|^4\\
+&\begin{cases}
g_{3}|m|^{3}cos3\theta, & \text{for $1/3$- and $2/3$-filling}\\
g_{6}|m|^{6}cos6\theta, & \text{for $1/2$-filling}
\end{cases}
 	\label{Eq6}
\end{aligned}
\end{equation}
where $|m|$ and $\theta$ represent modulus and phase of the order parameter in Eq.~\ref{Eq2}, $g_{2,3,4,6}$ are the coefficients of each term, and $|m|^{3} \cos3\theta$ and $|m|^{6} \cos6\theta$ are known as the three- and six-fold anisotropic terms breaking the $U(1)$ symmetry of $m$ down to $Z_3$ and $Z_6$\cite{r152,r155}, respectively. For the TAF ordered to disorder transition in a two-dimensional system at finite temperature, $Z_3$ anisotropic term is relevant, resulting in the phase transition belonging to the three-state Potts universality class\cite{r153}. However, $Z_6$ anisotropy is presumably irrelevant in the presence of strong thermal fluctuation, yielding an intermediate phase with emergent $U(1)$
isotropy of TAF order parameter, namely a KT phase\cite{Moessner2000PRB,r171}. The phase transition between the quasi long-range KT phase and the disordered phase at higher temperature belongs to the celebrated KT universality class\cite{r158,r156}. Thus, to further elaborate on the nature of TAF ordered phases at $\frac{1}{3}$/$\frac{2}{3}$- and $\frac{1}{2}$-filling, we embark on investigating the SSB and associated phase transitions at finite temperature.

\begin{figure}[t]
	\centerline{\includegraphics[scale=0.13]{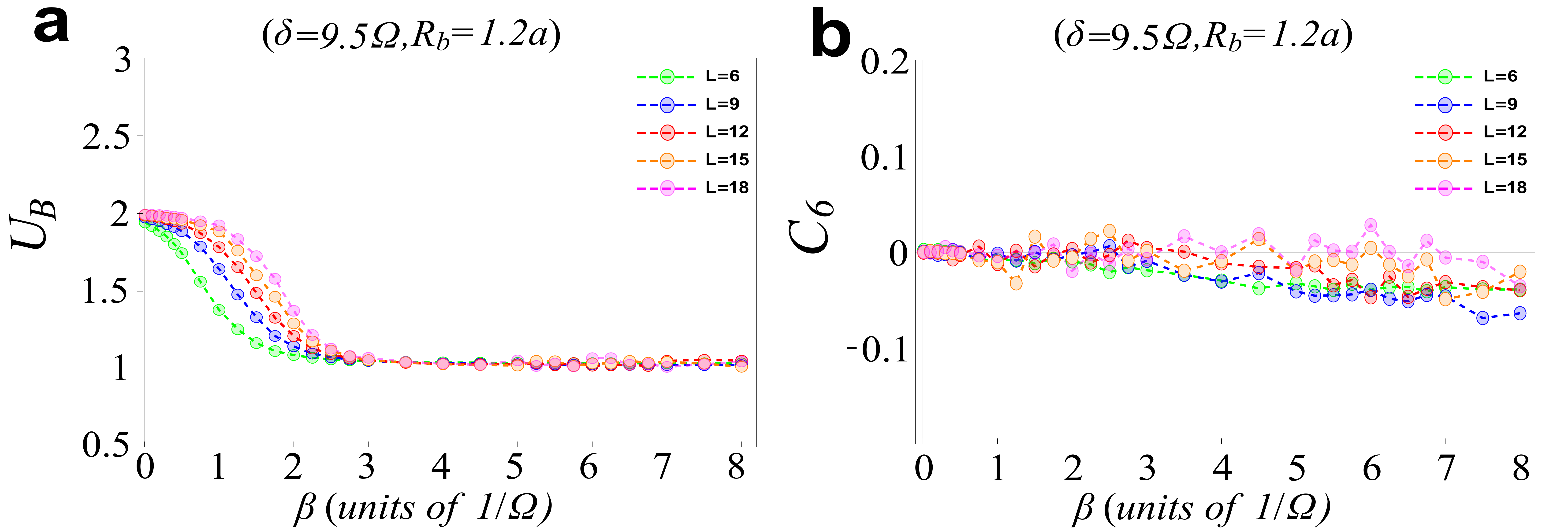}}
	\caption{{\bf Irrelevance of sixfold anisotropy.} (a) and (b) represent Binder ratio and $C_{6}$ for parameters $\delta=9.5\Omega$ corresponding to the $1/2$-filling of Rydberg occupancy. The system sizes in the simulation are $L=6,9,12,15,18$.}
	\label{Fig4}
\end{figure}

To explicitly characterize the anisotropy in the $\sqrt{3}\times\sqrt{3}$ TAF phases, we calculate the following quantity:
\begin{equation}
\begin{aligned}
C_{q}=\frac{\left< |m|^{q}cosq\theta \right>}{\left< |m|^{q} \right>},~~~q=3,6,
 	\label{Eq7}
\end{aligned}
\end{equation}
which is finite in the phase featuring three- or six-fold anisotropy\cite{r153,r155}. If the $Z_q$ anisotropic term is irrelevant, namely the ground-state manifold is a $U(1)$ group, such quantity is expected to vanish at thermodynamic limit. In addition, we also calculate Binder ratio to identify transition points at finite temperature, which are presented together with $C_{3}$ in Fig.\ref{Fig3}. The crossing feature of Binder ratio with varying system sizes in Fig.\ref{Fig3} (a) and (c) clearly demonstrate a finite-temperature transition from TAF phase to disordered phase, where the critical temperature at $\frac{1}{3}$- and $\frac{2}{3}$-filling are $\beta_c \Omega= 1.28\pm0.02$ and $1.22\pm0.02$, respectively. In the TAF phase at $\beta>\beta_c$ shown in Fig.~\ref{Fig3} (b) and (d), $|C_{3}|$ increases with system size, revealing the nature of anisotropy breaking $U(1)$ symmetry at $\frac{1}{3}$- and $\frac{2}{3}$-filling. The transition points identified from the crossing point of $C_{3}$ for different systems sizes are nearly consistent with the value determined by Binder ratio\cite{rSM2}. Moreover, we perform a systematic finite-size scaling analysis in SM\cite{rSM3} yielding the accurate critical exponents of the transition, which convincingly verifies the transition in Fig.~\ref{Fig3} belongs to the Potts universality class. In contrast, at $\frac{1}{2}$-filling, six-fold anisotropy $C_{6}$ vanishes with increasing system size in a large temperature regime indicated in Fig.~\ref{Fig4} (b), which implies the emergence of $U(1)$ symmetry. In the results of Binder ratio in Fig.~\ref{Fig4} (a), the crossing point is not explicitly detected, consistent with the absence of long-range order breaking continuous phase transition at finite temperature in two-dimension. The irrelevance of $Z_6$ anisotropy persists down to the lowest temperature accessible in our simulation $\beta=150/\Omega$.

\begin{figure}[t]
	\centerline{\includegraphics[scale=0.22]{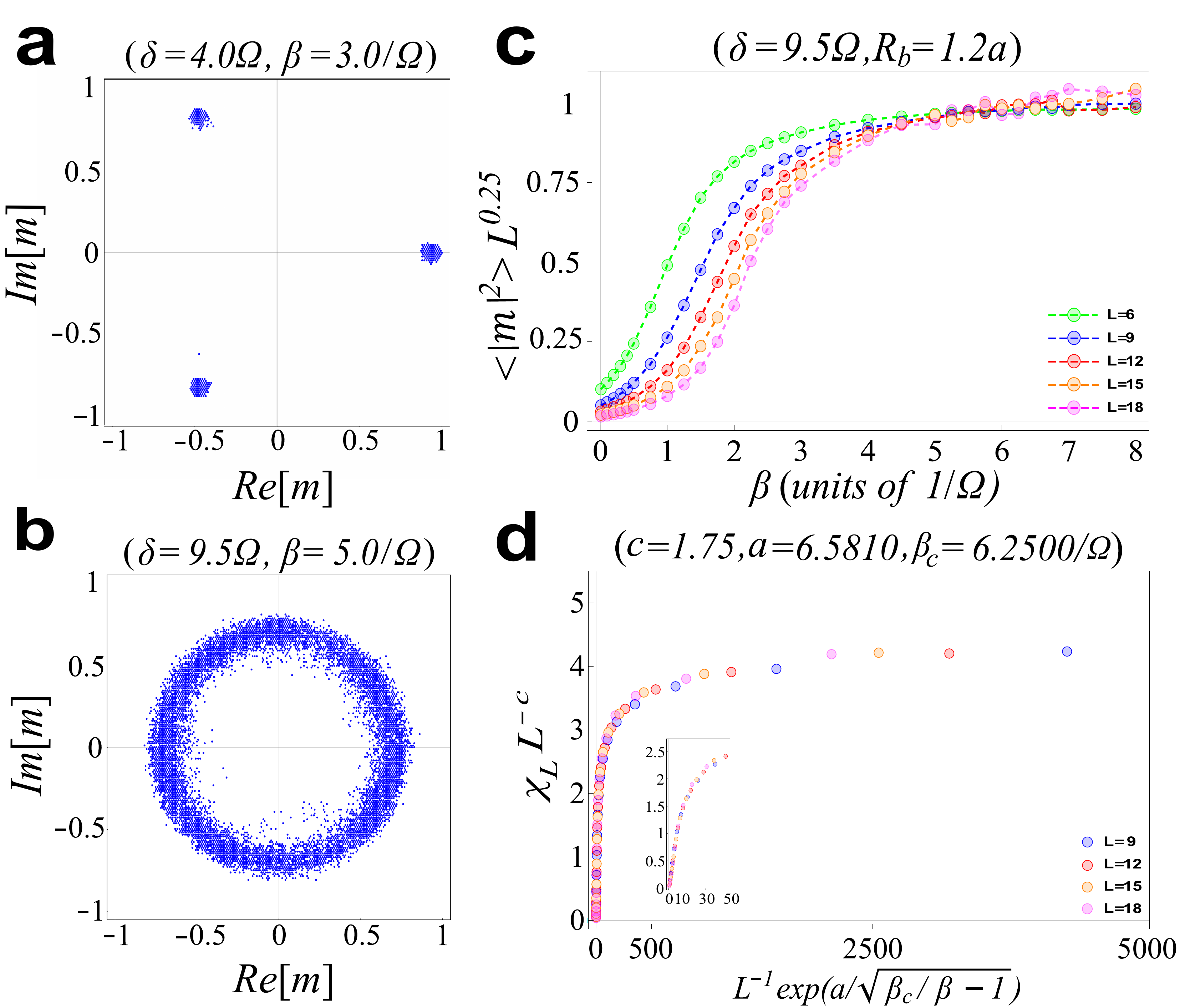}}
	\caption{ {\bf Histogram of the order parameter and data collapse of the susceptibility.} (a) and (b) represent the distribution histograms of TAF order parameters at $\frac{1}{3}$- and $\frac{1}{2}$-filling of Rydberg occupancy, respectively.  (c) The scaled TAF structure factor as a function of $\beta$, where the crossing point for different $L$ indicates the KT-transition point. (d) show the finite-size scaling analysis of TAF susceptibility, where the inset in (d) shows the fitting results when $\xi/L \in [0,50]$.}
	\label{Fig5}
\end{figure}

To further confirm the emergence of $U(1)$ symmetry in the TAF phase arising from OBD mechanism at $\frac{1}{2}$-filling, we plot the distribution histograms of TAF order parameters in \Fig{Fig5}(a)-(b). \Fig{Fig5}(a) is the histogram of $m$ at $\frac{1}{3}$-filling exhibiting pronounced three-fold anisotropy. In contrast, as depicted in \Fig{Fig5}(b),  the distribution histogram of TAF order parameter at $\frac{1}{2}$-filling is isotropic in the complex plane, providing unequivocal evidence of emergent $U(1)$ symmetry. Remarkably, even at $\beta=150/\Omega$ corresponding to temperature $T=70nK$ in Rydberg atom array experiment\cite{Browaeys2021Nature}, our calculation in SM\cite{rSM3} reveals that the $U(1)$ symmetry remains present in the histogram of $m$ at $\frac{1}{2}$-filling, manifesting the irrelevance of $Z_6$ anisotropy at such low temperature.

Furthermore, we decipher the nature of $U(1)$ symmetric TAF phase by investigating the finite-temperature phase transition to the high-temperature disordered phase, belonging to the celebrated KT-transition universality class owing to the emergent continuous $U(1)$ symmetry. Close to the KT-transition point, the susceptibilities of the $U(1)$ order parameters at various system sizes obey the scaling function\cite{KTpaper,r153,r155}:
\begin{equation}
\begin{aligned}
\chi_{L} =L^{2-\eta}\chi_{0}(\xi /L),
 	\label{Eq9}
\end{aligned}
\end{equation}
where $\chi_{0}$ is an unknown scaling function, $\xi\propto e^{a/\sqrt{\beta_{c} /\beta-1}}$ represents the correlation length of order parameters, displaying divergence at KT-transition inverse temperature $\beta_c$, and $a$ is a non-universal constant. The anomalous dimension of order parameter $\eta=0.25$ exactly holds in KT-transition\cite{r158,r156,r153,Moessner2000PRB}. We present the results of rescaled susceptibility $\left< |m|^{2} \right> L^{0.25}$ as a function of inverse temperature $\beta$ in \Fig{Fig5}(c), which can be deduced from the order parameter in our simulation through the relation: $\chi =L^{2}\left< |m|^{2} \right> \beta$, clearly indicating the KT transition occurs at $\beta_c = 6.25/\Omega$ by virtue of the crossing point of rescaled susceptibilities for different system sizes.

Next, we adopt data collapse analysis of the TAF susceptibility using the scaling function \Eq{Eq9}. Assuming KT transition universality class, the data points $(L^{-1}e^{a/\sqrt{\beta_{c} /\beta-1}}, \chi_L L^{-1.75})$ should collapse into a single curve with appropriate choice of parameters\cite{r153,r155}. Employing the approach of data collapse analysis\cite{r154,r157}, we determine the phase transition temperature $\beta_c\Omega=6.2513\pm 0.9082$ and $a=6.5842\pm 0.9975$, and all data points of each group collapse into a single smooth curve as shown in Fig.~\ref{Fig5} (d). The estimated value of $\beta_c$ from data collapse is consistent with the result obtained in \Fig{Fig5}(c) within error bar. More crucially, the results of data collapse provide convincing evidence that the phase transition between critical KT phase and disordered phases belongs to KT transition, further verifying the irrelevance of six-fold anisotropy at half filling with increasing thermal fluctuation.

{\em Conclusions and discussions.}---Employing numerically exact QMC simulation, we systematically investigate a realistic interacting model describing triangular Rydberg atom array. Upon fixing Rydberg radius $R_b$, the ground state phase diagram with varying Rydberg filling is achieved. At $\frac{1}{3}$- and $\frac{2}{3}$-filling, the $\sqrt{3}\times\sqrt{3}$ TAF long-range order is unambiguously unveiled, consistent with the observation in recent Rydberg atom array experiments on triangular lattice\cite{Browaeys2021Nature}. More appealingly, our calculation reveals the $\sqrt{3}\times\sqrt{3}$ ordering is present at $\frac{1}{2}$-filling, arising from OBD mechanism. Owing to the interplay between quantum and thermal fluctuations, an enlarged $U(1)$ isotropy of TAF order parameter is emerged in a large temperature regime, resulting in a KT transition between the TAF and disordered phases. Since the classical patterns of TAF order at $\frac{1}{3}$- and $\frac{2}{3}$-filling have been successfully imaged in recent experiment\cite{Browaeys2021Nature}, our state-of-the-art numerical simulation paves the route for subsequent observations of quantum fluctuation triggered OBD $\sqrt{3}\times\sqrt{3}$ TAF phase and the intriguing physics including emergent KT phase and phase transition in Rydberg atom platform.

In addition, it is fascinating to investigate the model \Eq{Eq1} with further increasing Rydberg blocking radius  $R_{b}$. As discussed in previous studies, various magnetic ordered phases\cite{r172,r173}, more intriguingly, quantum spin liquid phase featuring topological order could possibly emerge by varying Rydberg blocking radius\cite{r173,Lukin2021Science}. Moreover, it is straightforward to generalize our study to Rydberg systems in other lattices or modified interactions. Hence, we believe that our study opens a new avenue to investigating exotic physics emerging in the Rydberg atom array on frustrated lattice by unbiased numerical simulation.

{\em Acknowledgments.}----S.G. and J.H. is supported by the Ministry of Science and Technology  (Grant No. 2022YFA1403901), the National Natural Science Foundation of China (Grant No. NSFC-11888101),  the New Cornerstone Investigator Program and the Strategic Priority Research Program of Chinese Academy of Sciences (Grant No. XDB28000000). Z.X.L acknowledges support from the start-up grant of IOP-CAS.

\bibliography{ref}
\bibliographystyle{apsrev4-1}

\end{document}